\documentclass{llncs}
\usepackage{amsmath}
\usepackage{amssymb}
\usepackage[all]{xy}
\usepackage{graphicx}
\usepackage{url}

\newcommand{\equprogram}[1]{
\def\separator{1.5ex}
\frenchspacing
\refstepcounter{equation}%
\par\vspace\separator\hspace{2.5em}%
$\vcenter{\tt\noindent\kern-.5em{#1}}$%
\kern-45pt\llap{(\theequation)}
\par\vspace\separator\noindent\kern-.0em%
}

\newcommand{\trs}{\ensuremath{\mathcal{R}}}
\newcommand{\vars}{\ensuremath{\mathcal{X}}}
\newcommand{\constrs}{\ensuremath{\mathcal{C}}}
\newcommand{\opers}{\ensuremath{\mathcal{D}}}

\newcommand{\bool}{\ensuremath{\mathbb{B}}}

\begin{document}

\title{An Implementation of Bubbling}

\author{
Abdulla Alqaddoumi
\quad Enrico Pontelli
}

\institute{
Department of Computer Science \\
New Mexico State University \\
Las Cruces, NM 88003, U.S.A.\\
\email{aalqaddo,epontell@cs.nmsu.edu} 
}
\maketitle
\begin{abstract}
{Non-determinism is of great importance in functional logic programming. It provides expressiveness and efficiency to functional logic computations. In this paper we describe an implementation of the multi-paradigm functional logic language Curry. The evaluation strategy employed by the implementation is based on definitional trees and needed narrowing for deterministic operations, while non-deterministic operations will depend on the graph transformation, bubbling. Bubbling preserves the completeness of non-deterministic operations and avoids unnecessary large-scale reconstruction of expressions done by other approaches.} 
\end{abstract}
\section{Introduction}\label{introductionSect}

{Non-determinism \cite{Antoy98optimalnon-deterministic} is of great importance to functional logic computations. Functional logic programs using non-determinism can be more expressive than functional programs. Non-determinism allow computations to possibly yield more than one value.}   

{Functional logic programming \cite{AntoyHanus10CACM,Bellia86Relation,Hanus07ICLP} is a multi\--paradigm programming that combines in a seamless way the best features of functional programming and logic programming. Functional programming is based on $\lambda$-calculus and provides the users with features like demand driven evaluation, polymorphic typing and higher-order functions. Logic programming is based on Horn clause logic, a subset of first order logic. Logic programming provides reasoning with partial data and missing information, non-determinism, logical variables, and function inversion. Therefore, functional logic languages have several advantages over functional and logic programming. Functional logic languages are more expressive than functional languages thanks to non-determinism, logical variables, and the ability to deal with infinite data structures. Functional logic languages are more efficient than logic languages because of demand-driven evaluation.}

{Current implementations of functional logic languages belong to one of three categories: (1) Implementations that include the logic programming features in a functional language. (2) Implementations that extend logic languages with functional programming features. (3) Implementations of virtual machines that implement the features of logic programming and functional programming using an imperative language. Interested readers are referred to \cite{Hanus07ICLP,AntoyHanus10CACM} for a survey of such languages and their implementations.}

{In this paper, we present a design for an implementation of a virtual machine for the functional logic language Curry \cite{Hanus06Curry}, a community-agreed standard language. The implementation will focus on issues concerning non-determinism and logical variables that other implementations, we believe, do not tackle correctly. Bubbling \cite{Antoy06Correctness,lazycontext} is a graph transformation that correctly tackles such concerns.}

{In section \ref{backgroundSect} we briefly review some concepts of functional logic programming. In section \ref{BubblingSect} we will discuss the key concepts of bubbling in details. In section \ref{implementationDetailsSect} we will describe the current implementation. In section \ref{FutureWorkSect} we discuss future work and conclude in section \ref{conclusionSect}.}

\section{Background}\label{backgroundSect}

\subsection{Preliminaries}\label{PreliminariesSubSect}

{Functional logic programs can be modeled as a class of constructor-based conditional term graph rewriting systems (TGRS). Terms are modeled as graphs to allow sharing of sub-terms.}
\newline
{We recall the definition of term rewriting systems from \cite{termrewriting}.}

{A rewrite system is a pair, \trs = $\langle \Sigma$, R$\rangle$, where $\Sigma$ is a signature and R is a set of rewrite rules. The signature $\Sigma$ consists of different symbols where each symbol is associated with an arity and an operator designation. The main operator designations are $constructors$ and $defined$ $operations$. $Constructors$, \constrs, are specific operators that are used to construct data, whereas, $defined$ $operations$, \opers, operate on data to transform it into a $constructor$ $term$. Numbers and list constructors, such as ``cons'' and ``nil'', are examples of $constructors$, while multiplication ``$\times$'' and division ``/'' are examples of $defined$ $operations$. $Constructors$ with no arguments (0-ary) are usually referred to as $constants$. Variables are present in functional logic programming and have a different meaning from their counterparts in imperative programming. Variables in functional logic programming can be assigned a value at most once in the whole execution and have no arguments. Variables belong to the countably infinite set \vars. The set of terms, T($\Sigma$, \vars), is defined inductively as follows: (1) All variables x $\subseteq$ T($\Sigma$, \vars). (2) For all f $\in \Sigma$ of arity n $\geq$ 0 and $t_1$, $\ldots$, $t_n$ $\in$ T($\Sigma$, \vars), f($t_1$, $\ldots$, $t_n$) $\in$ T($\Sigma$, \vars). A term, f($t_1$, $t_2$, $\ldots$, $t_n$), is called $constructor$-$rooted$ or $operation$-$rooted$ if the operator f is a $constructor$ or $defined$ $operation$, respectively. A term, f($t_1$, $t_2$, $\ldots$, $t_n$), is called a $pattern$ if the operator f is a defined operation and $t_1$, $t_2$, $\ldots$, $t_n$ are $constructor$-$rooted$ terms. Variables and $constructor$-$rooted$ terms represent values commonly known as $head$ $normal$ $form$. Var(t) refers to the set of variables in the term t. A term t is said to be $linear$ if each variable appears in it at most once. A term t is said to be ground if Var(t) = $\emptyset$. Access to sub-terms is done through the operator $|$ followed by a (possibly empty) sequence of natural numbers. For example, in the term f($t_1$, $t_2$, $\ldots$, $t_n$), access to the sub-term $t_p$ for 1 $\leq$ p $\leq$ n is denoted by f($t_1$, $t_2$, $\ldots$, $t_n)|_p$. The access to the root is done through f($t_1$, $t_2$, $\ldots$, $t_n)|_\varepsilon$. The replacement of a sub-term in term t at position p, $t|_p$, with another term s is denoted by $t[s]_p$. For example, f($t_1$, $t_2$, $\ldots$, $t_n)[s]|_2$ corresponds to the term f($t_1$, s, $\ldots$, $t_n$).}

{Rewrite rules, R, are of the form l {\bf =} r where l and r are terms; we refer to l (r) as the left-hand-side (right-hand-side) of the rule. A rewrite system \trs \hspace*{1pt}  is of type left-linear if all the left-hand-sides of the rules are linear. \trs \hspace*{1pt} is said to be constructor-based if all left-hand-sides of the rules are patterns.}

{A substitution is a mapping $\sigma$ : \vars $\to$ T($\Sigma$, \vars) that maps variables to terms. The application of $\sigma$ to variables in a term t, $\sigma$(t), replaces the variables that appear in t with their value in $\sigma$. Two terms t and s are unified if there exists a substitution $\sigma$ such that $\sigma$(t) = $\sigma$(s). A rewriting step t $\rightarrow _{p, R}$ s is done when there exists a position p in term t, a rewrite rule l = r in R and a substitution $\sigma$ such that $t|_p$ = $\sigma$(l), and s = t[$\sigma$(r)$]_p$. A term t is said to be $irreducible$ or in $normal$ $form$ if t cannot be rewritten to any other term using the rules in $R$.}

{We will use graphs to represent terms. Each graph consists of a set of nodes and edges connecting these nodes. The nodes are labeled with an operator designation or a variable. In the following, we extend the definition of term graphs with a single root, commonly known as expressions, from \cite{EchahedGraph} with some nominal changes. Let $\Sigma$ be a signature, \vars \hspace*{1pt} a countable set of variables, and N a countable set of nodes.}

{A rooted graph over $<\Sigma$, N, \vars$>$ is a 4-tuple GRAPH g = $\langle N_g$, $L_g$, $E_g$, $Root_g \rangle$ such that:}
\begin{enumerate}
\item	$N_g \subseteq$ N is the set of nodes of g.
\item	$L_g$: $N_g$ $\rightarrow \Sigma \cup$\vars \hspace*{1pt} is the labeling function that maps each node of g to a signature symbol or a variable.
\item	$E_g$: $N_g$ $\to$ Nat $\to$ $N_g$ is the edge function mapping each node of g to the node that represent its sub-expressions. For a node n in g where $L_g$(n) = s, s $\in \Sigma \cup$\vars, and arity(s) = k then $E_g$(n, 1) = $n_1$ ,$\ldots$, $E_g$ (n, k) = $n_k$, $n_1, \ldots, n_k$ $\in$ $N_g$.
\item	$Root_g$ $\in$ $N_g$ is a node in g that is called the root of g.
\item	Every variable in the graph g labels one and only one node. That is, if $L_g$($n_1$) $\in$ \vars \hspace*{1pt} and $L_g$($n_2$) $\in$ \vars, then $L_g$($n_1$) = $L_g$($n_2$) $\Longrightarrow$ $n_1$ = $n_2$.
\item	Each node in g is either $Root_g$ or reachable from $Root_g$ through the edge function $E_g$.
\end{enumerate}

\equprogram{
\label{graphExEq}
(3 + 4) - 5
}

\begin{figure}[h!]
\centering
\includegraphics[scale=0.2] {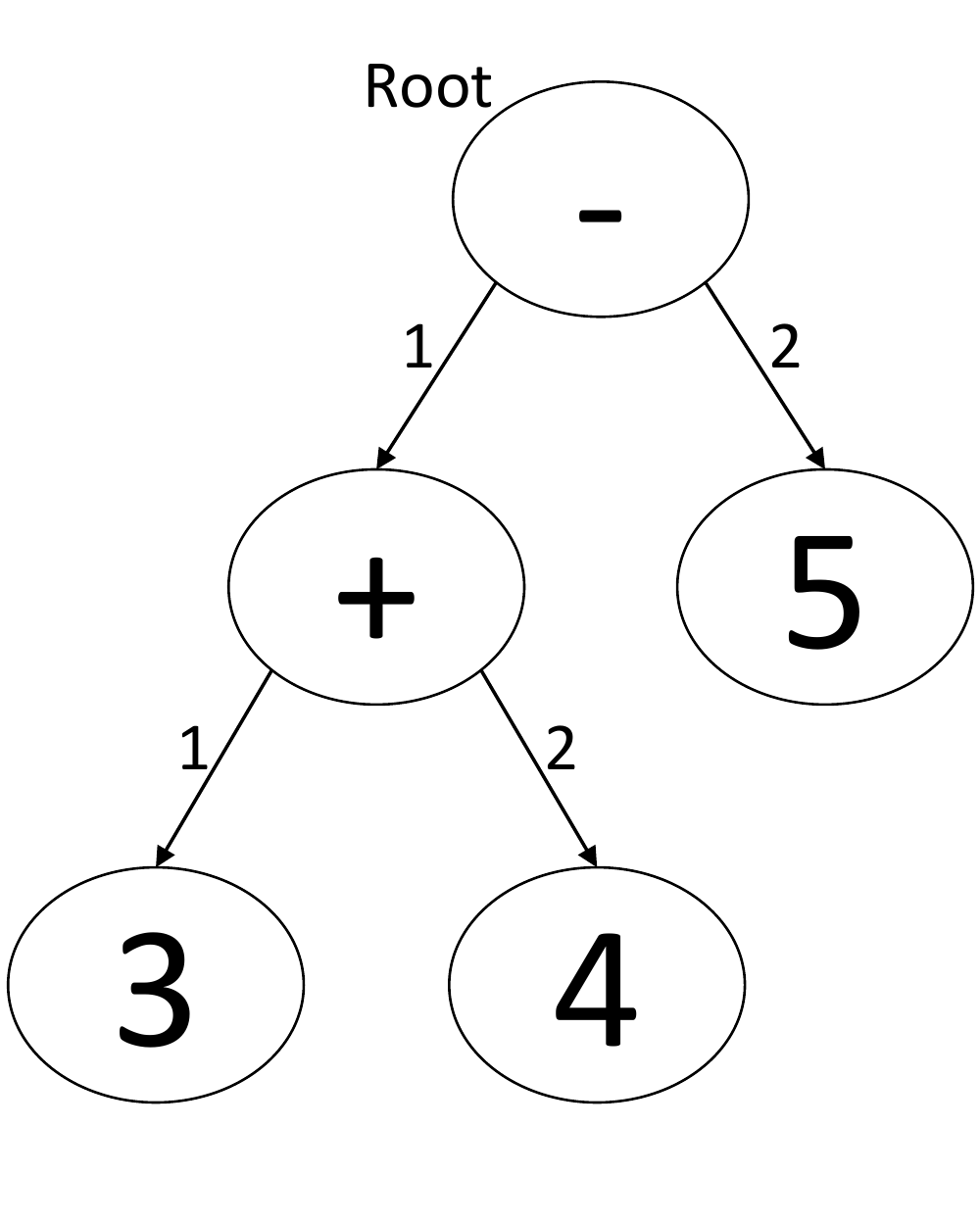}
\caption{Pictorial representation for the expression in equation \ref{graphExEq}}
\label{fig:Graphexample}
\end{figure}

Figure \ref{fig:Graphexample} is a pictorial representation of the expression in equation \ref{graphExEq}. The symbols The labels of the nodes, $L_g$, represent the symbol  of the node. The root of the expression is the node with the name ``Root''. The edge function is represented by the arrows that connect the nodes.

{A term rewrite system specifies the rewrite rules but does not specify the precedence of rules over each other. Also it does not specify which arguments must be evaluated and in which order. All these details are left for an evaluation strategy that controls and guides the rewriting process. Antoy introduced in \cite{defTrees} a hierarchy, the $definitional$ $trees$, to order all the rewrite rules and to specify which arguments should be selected. A definitional tree for a defined operation f is a set of linear patterns partially ordered by subsumption. The leaves of the tree are variants of the left-hand sides of the rules defining f. The root is a pattern of the form f($X_1$, \ldots, $X_n$) where $X_1$, \ldots ,$X_n$ are distinct variables. The inner nodes have an $inductive$ $position$, where the children of such inner node will have different constructor terms.}

{The following are the rewrite rules that define the operation less than or equal ``$\leq$'' \cite{defTrees} using Peano numbers and boolean values. Note that ``$\_$'' represents an anonymous variable. An anonymous variable does not require any computation for its value.}
\equprogram{
\label{lesseqOP}
0 $\leq$ $\_$ = true \newline
s($\_$) $\leq$ 0 = false \newline
s(X) $\leq$ s(Y) = X $\leq$ Y
}
\begin{figure}[ht]
\centering
\includegraphics[scale=0.17]{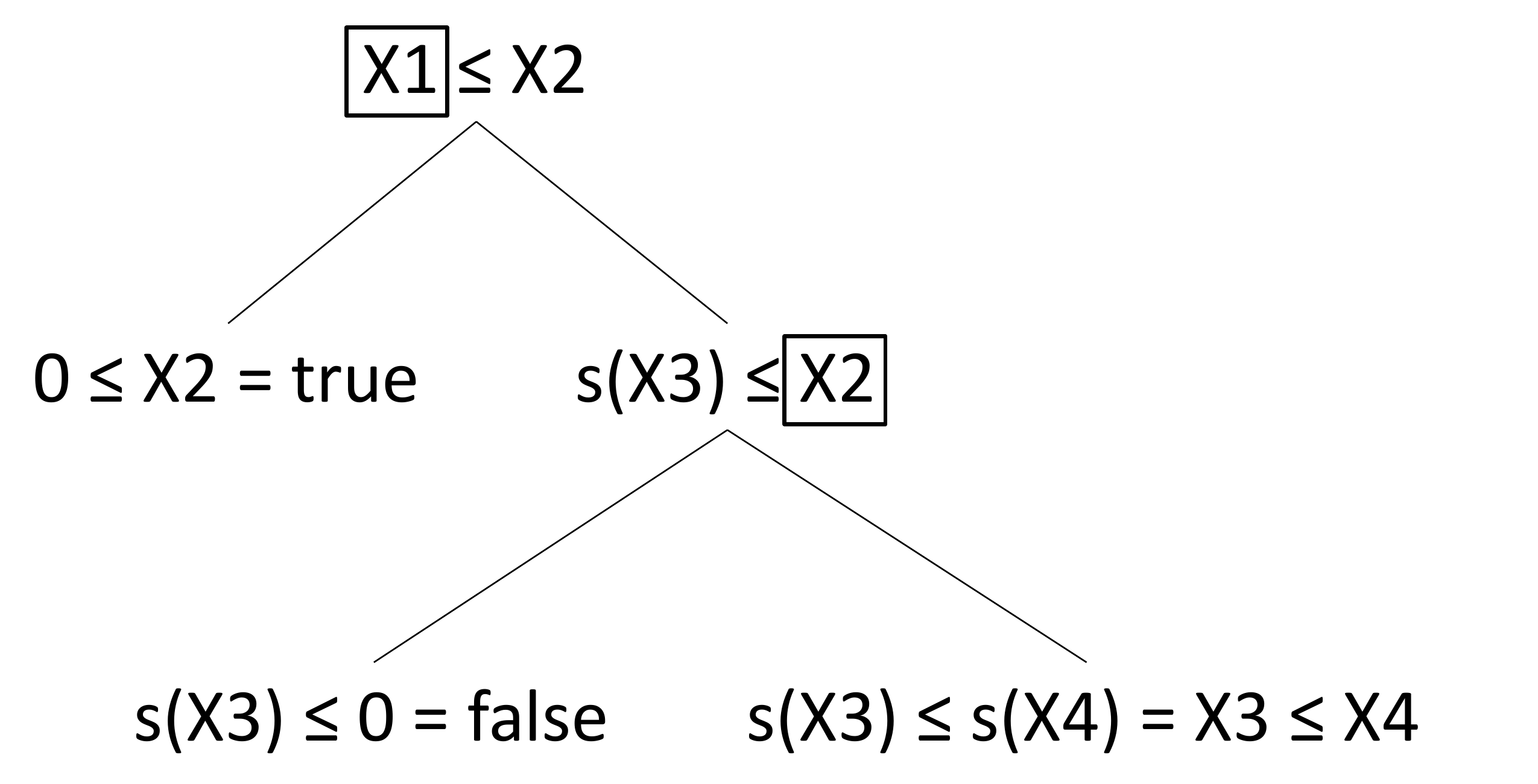}
\caption{Pictorial representation of the definitinal tree for the defined operation $\leq$ appearing in equation \ref{lesseqOP}. The framed variables represent the position needing evaluation.}
\label{fig:2}
\end{figure}

{Figure \ref{fig:2} is a definitional tree for the operation $\leq$, where the inductive positions are surrounded by a box. The root of the tree has two variables X1 and X2. The children of the root differ in the instantiation of the variable X1. When X1 is instantiated to 0, X2 does not need further evaluation and the expression will be rewritten to {\bf true}. When X1 is instantiated to s(X3), the tree will instantiate X2 to either 0 or s(X4). In the first case, the expression will be evaluated to {\bf false}, while the later will require a re-application of the first rule (root) with the fresh variables X3 and X4. Fresh variables are variables that never appeared before. For more details about definitional trees, please refer to \cite{defTrees}.}

\subsection{Non-determinism}\label{NonDeterminismSubSect2}

{\indent\indent Non-deterministic operations and logical variables \cite{Hanus94Survey,Antoy06overlapping} play an important role in functional logic programming. Non-determinism increases the expressiveness of functional logic programs.}

{The choice operator, ``?'', is used to express non-determinism. Any defined operation that is rewritten to an expression that uses a choice operator is considered a non-deterministic operation.}
\equprogram{
\label{choiceOp}
x ? y {\bf =} x \\
x ? y {\bf =} y}

{Rules in equation (\ref{choiceOp}) define the choice operator used in a rewrite rule. One of the arguments of the choice will be chosen non-deterministically to rewrite the left-hand-side of the rule.}

{We recall two classes of constructor-based term rewriting systems relevant to our paper, $inductively$ $sequential$ and $overlapping$ $inductively$ $sequential$ classes. The $inductively$ $sequential$ class (IS) represents the rules that constitute the first-order component of any functional language, and hence functional logic languages as well. A defined operation $f$ is called $inductively$ $sequential$ if there is a definitional tree of $f$ that contain all and the only rules defining $f$. A TRS is of type $inductively$ $sequential$ if all its defined operations are inductively sequential operations. The $overlapping$ $inductively$ $sequential$ class (OIS) is a super class of the IS class. The OIS class adds non-deterministic rules to the set of rules that are $inductively$ $sequential$. The term rewriting system class that will be used to model functional logic programs in this implementation is the OIS class \cite{Antoy01evaluationstrategies}.}

{Modeling functional logic programs in the OIS class is not a limitation as shown in \cite{Antoy01ConsCond}. Antoy introduced a set of transformations that will convert a constructor-based conditional TRS into one of its sub-classes, the $overlapping$ $inductively$ $sequential$ TRS. We will show in section \ref{BubblingSect} an efficient evaluation strategy to evaluate programs modeled by the OIS class.}

{The presence of variables in expressions represent that: (1) An expression is shared between two or more expressions. (2) A logical variable. Logical variables express non-determinism. Antoy et al. introduced in \cite{Antoy06overlapping} a transformation that eliminates logical variables by replacing the logical variables with overlapping rules.} 

\equprogram{
\label{sharingEX}
(1+X)+(X+2) where X = (0 ? 1)
}

\begin{figure}%
\centering
\begin{minipage}[ht]{0.43\linewidth}
\includegraphics[scale=0.2]{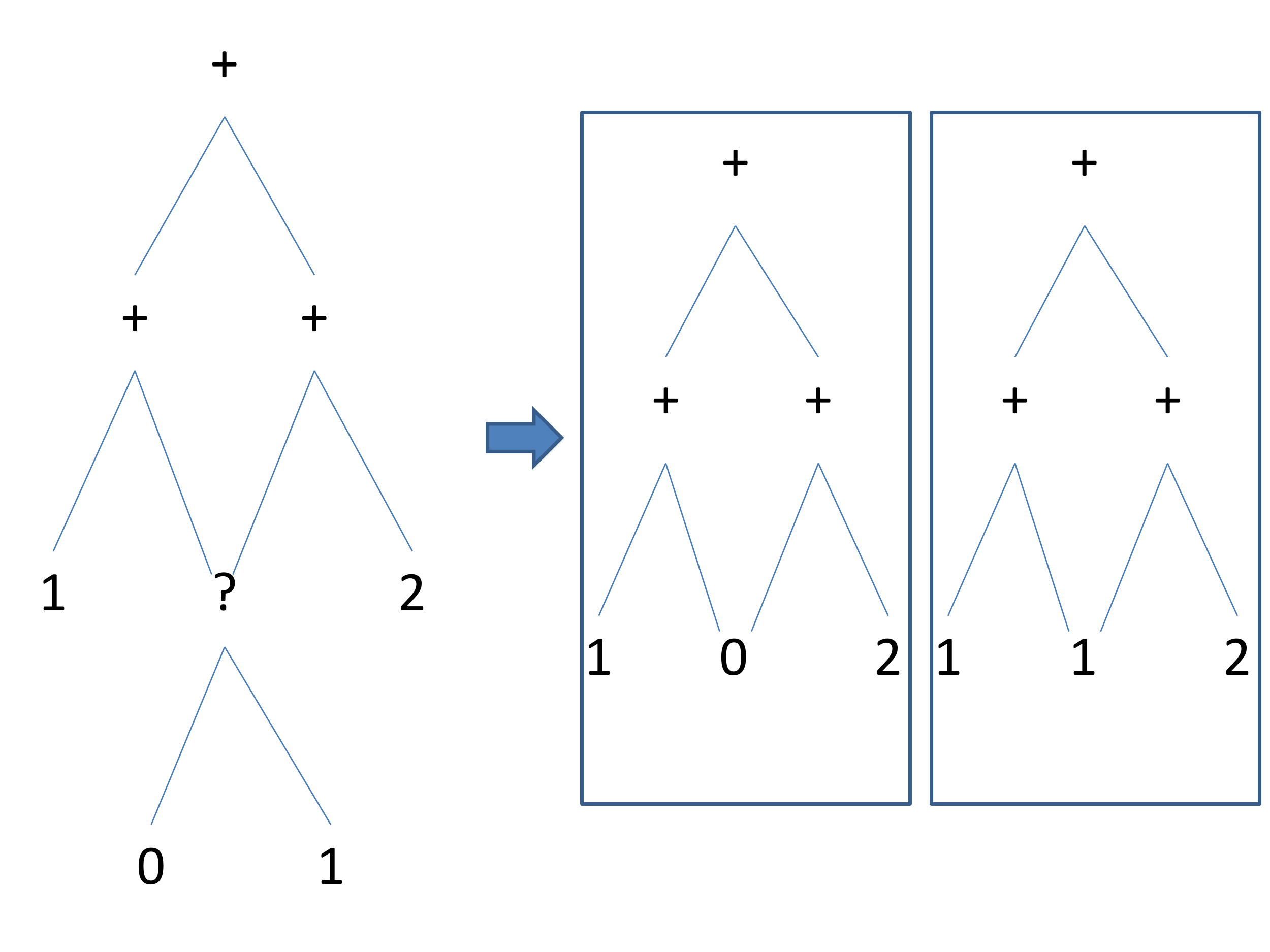}
\caption{Pictorial representation of equation \ref{sharingEX} (sharing)}
\label{fig:2figsA}
\end{minipage}
\qquad
\begin{minipage}[ht]{0.43\linewidth}
\includegraphics[scale=0.2]{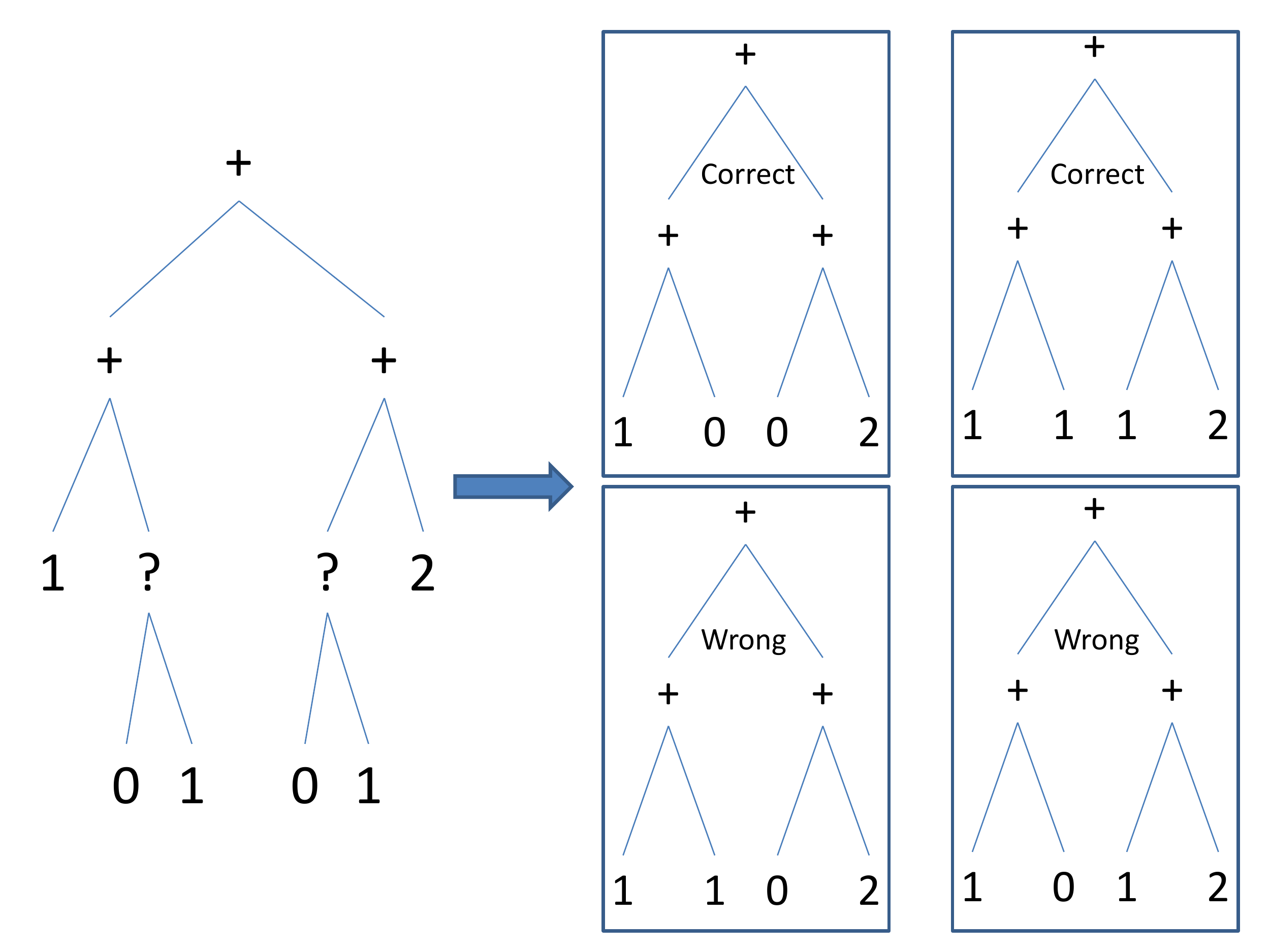}
\caption{Pictorial representation of equation \ref{sharingEX} (duplication)}
\label{fig:2figsB}
\end{minipage}
\end{figure}

{The expression in equation \ref{sharingEX} represents sharing. Sharing is introduced in expressions through the use of the ``where'' clause. If two or more expressions require the evaluation of a shared sub-expression, then the shared sub-expression will be evaluated only once. If the shared sub-expression is duplicated, then the evaluation of the shared sub-expression will be done more than once. In the case that the shared sub-expression is a deterministic operation, the multiple evaluations will not affect the outcome of the computation, but will affect its efficiency. On the other hand, if the shared sub-expression is a non-deterministic operation, then the outcome of the computation will not be correct. The following figures, \ref{fig:2figsA} and \ref{fig:2figsB}, represent the expression in equation \ref{sharingEX}. The first one supports sharing and the second one converts sharing into duplication of shared expression. The rewriting of ``?'' produces the two only correct expressions in figure \ref{fig:2figsA}, while the rewriting of ``?'' in figure \ref{fig:2figsB} produces the two correct expressions along with two wrong ones.}

\subsection{Evaluation strategy}\label{EvalStrategySubSect}

{\indent\indent Evaluation of a logic programming goal is done by the construction of a proof in SLD-resolution. When a goal contains variables, then the system will provide the user with possible bindings for such variables that make the goal succeed. Evaluation in functional programming is done by the repeated rewriting of an expression until a $normal$ $form$ or an $irreducible$ expression is reached which is not a $normal$ $form$. The first case represents success if the $normal$ $form$ is a constructor term graph and all other cases represent failure of the computation.}

{Evaluation of functional logic expressions is a mixture of both these approaches. When we try to compute an expression that unifies with the left-hand-side of a deterministic rule, the expression will be rewritten to the right-hand-side of the same rule, similar to rewriting in functional programming. If the expression to be computed unifies with the left-hand-side of a non-deterministic rule, the expression can be rewritten to one of the right-hand-sides of the rule. Dealing with missing information is done by placing logical variables in the expressions. The instantiation of logical variables is done by narrowing. Narrowing is the process of guessing a $constructor$ $term$ to substitute for a variable appearing in an $operation$-$rooted$ expression. Thus, a logical variable can be rewritten to the non-deterministic choice operation with all guessed $constructor$ $terms$ becoming as arguments of such choice. Narrowing is comparable to unification and resolution in logic programming \cite{Bellia86Relation,Bosco88NarrowingSLD} and it is what essentially combines functional programming with logic programming.}

{The evaluation strategy used to compute programs in the IS class is Needed Narrowing \cite{NeededNarrowing}. When the evaluation strategy encounters an uninstantiated variable, then it will either narrow the variable or residuate. Narrowing will allow expressions with uninstantiated variables to proceed in the computation process without suspension. Narrowing is complete and computes all solutions possible. Residuation, on the other hand, delays and suspends the computation of sub-expressions containing the uninstantiated variable. The computation will continue only when such variable is instantiated while computing other sub-expressions. The absence of other sub-expressions that may compute values for such variable may cause residuation to suspend forever and, therefore, be unable to complete the computation. Needed narrowing uses narrowing to instantiate variables only when such variables are needed. The guidelines for needed narrowing depends on the definitional tree that contains the defined operation's rules. Non-deterministic operations belonging to the OIS class of term rewriting systems can be computed using the $Inductively$ $sequential$ $Narrowing$ Strategy, INS \cite{Antoy98optimalnon-deterministic}. INS is based on needed narrowing but lacks the ability to deal with sharing of sub-expressions.}

\subsection{Previous approaches}\label{PrevAppSubSect}
{Computing functional logic programs with non-determinism and sharing can be tackled using different approaches. Backtracking is used to compute the alternatives of a non-deterministic operation one by one. Whenever an alternative fails in computing a value, the next alternative is tried. This approach is used in implementations that transforms functional logic programs into Prolog programs, like PACKS \cite{Hanus08PAKCS}, an implementation of Curry.}

\equprogram{
\label{loopOP}
loop {\bf =} loop
}

\equprogram{
\label{loopEx}
(loop ? 1+2) 
}

{The computation of the expression appearing in equation \ref{loopEx} using backtracking may never produce a $normal$ $form$ if the argument of ``?'' selected is the loop operation, while the value 3 can be produced using other approaches. Such behavior is referred to as the problem of incompleteness.}

\equprogram{
\label{failEx}
f x {\bf =} 1+(2+($\ldots$+(100 / x)$\ldots$))
}

{Copying is an approach that is successful in tackling non-determinism and sharing. The concept is to reconstruct several copies of the whole expression for all the arguments of the non-deterministic operation ``?'' and to plug each argument of the ``?'' in one of the copies. Consider the computation of the expression (f (0?1)) using copying, where f is defined by the rule in equation \ref{failEx} (borrowed from \cite{lazycontext}). The expressions (f 0) and (f 1) will be built to compute the alternatives of the choice. The expression (f 0), rewritten as 1+(2+($\ldots$+(100 / 0)$\ldots$)), will fail immediately after the division by zero, (100 / 0), is encountered. All the effort in the construction of the expression (f 0) is wasted because failure is encountered at a very early stage of the computation \cite{lazycontext}. Copying has been used in some experimental implementations \cite{Antoy05avirtual,Tolmach04implementingfunctional}.}

\section{Bubbling}\label{BubblingSect}

{Bubbling is a graph transformation that preserves the completeness of non-deterministic operations without the need to construct unnecessary copies for the whole original graph. The transformation depends on finding a dominator of a needed choice and then ``bubbling'' the choice up to take the place of its dominator. The choice will become the immediate parent of the dominator and several copies of all the nodes that fall between the choice and its dominator, the Ancestral Path, will be made. All the edge links between the nodes that are part of the ancestral path and other nodes of the graph will be preserved.}

{We introduce some properties that hold for the term graph rewriting system of interest and some functions that will help in the bubbling process.}

{[Acyclic structure]	$\forall$ x, y $\in N_g$. If x is reachable from y, then y cannot be reachable from x. That is, no cycles are allowed in g.}
\newline
{The acyclic structure of the graph does not limit the capabilities of TRS. We recall that admissible graphs contain no cycles among the defined operations \cite{Echahed-ICLP98,EchahedGraph}. We use this property to ensure efficient execution of the function PathToRoot that collects all the paths that connects a certain node  of the graph to the root.}

{[Parent] Parent : $N_g$ $\times$ $N_g$ $\to$ \bool. A node p is a parent of a node x if p is related to x through the edge function $E_g$.}

{BackPointers: $B_g$: $N_g$ $\to$ $2^{N_g}$ is the Back pointers function that maps a node x of a graph g to the set of the parent nodes of x.}

{Path: $2^{N_g}$. A path is a sequence of nodes connected to each other through the BackPointers function starting from a node x to $Root_g$. An example of a path is $n_1n_2\ldots n_k$, where $n_1$ is the parent of x, $n_i$ is the parent of $n_{i-1}$ $\forall$ i, 1 \textless \hspace*{1pt} i $\le$ k, and $n_k$ is $Root_g$.}

{PathToRoot: $N_g \to 2^{2^{N_g}}$ is a function that takes a node x and returns the set of all paths that connect x to {$Root_g$}.}

{[Root] $\forall$ x $\in N_g$ $\setminus$ $\{Root_g\}$, $Root_g$ is a dominator of x.}

{AncestralPath: $2^{2^{N_g}} \to 2^{N_g} \times N_g$ is a function that takes all the paths connecting a node x to $Root_g$ and returns a node a, a dominator, and the set of nodes, AP, that forms the backbone that connects x to the dominator a.}

\equprogram{
\label{BubbleExampleEq}
(Fact X + Fibo X) where X = ( 2 ? 3 )
}

\begin{figure}
\centering
\begin{minipage}[ht]{0.4\linewidth}
\centering
\includegraphics[scale=0.17]{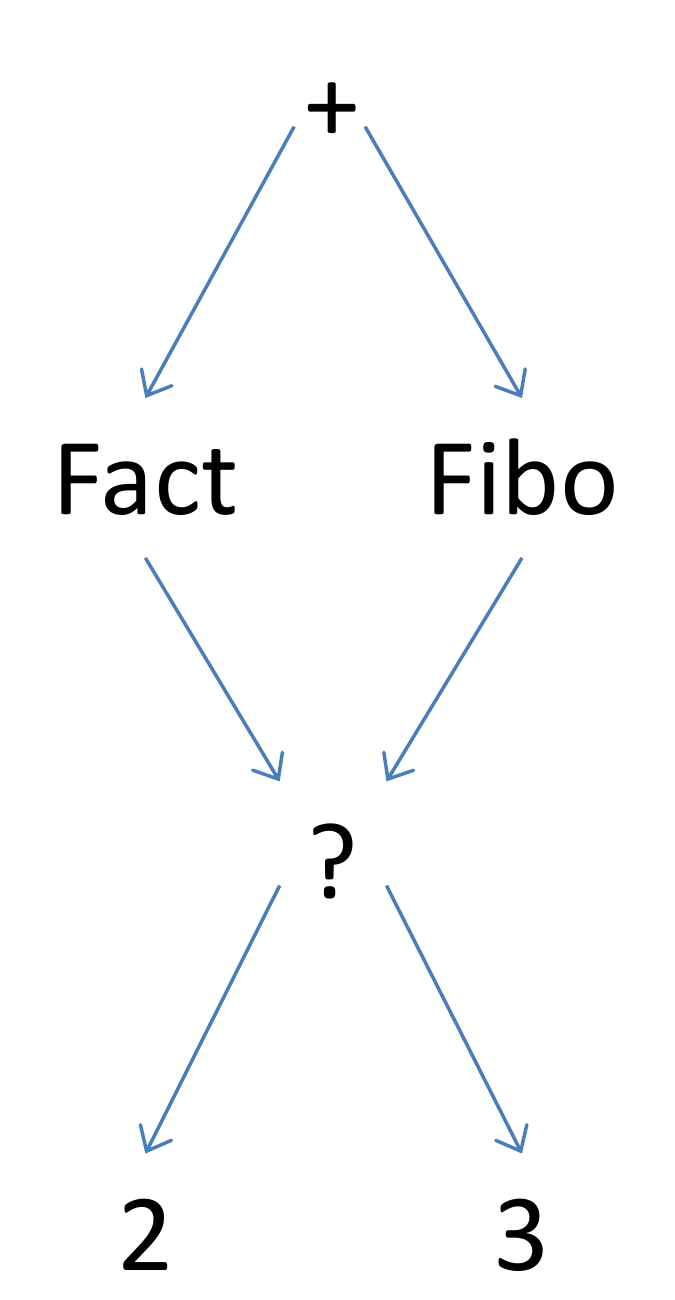}
\caption{Pictorial representation of the expression in equation \ref{BubbleExampleEq}}
\label{fig:BubblingExA}
\end{minipage}
\qquad
\begin{minipage}[ht]{0.4\linewidth}
\centering
\includegraphics[scale=0.17]{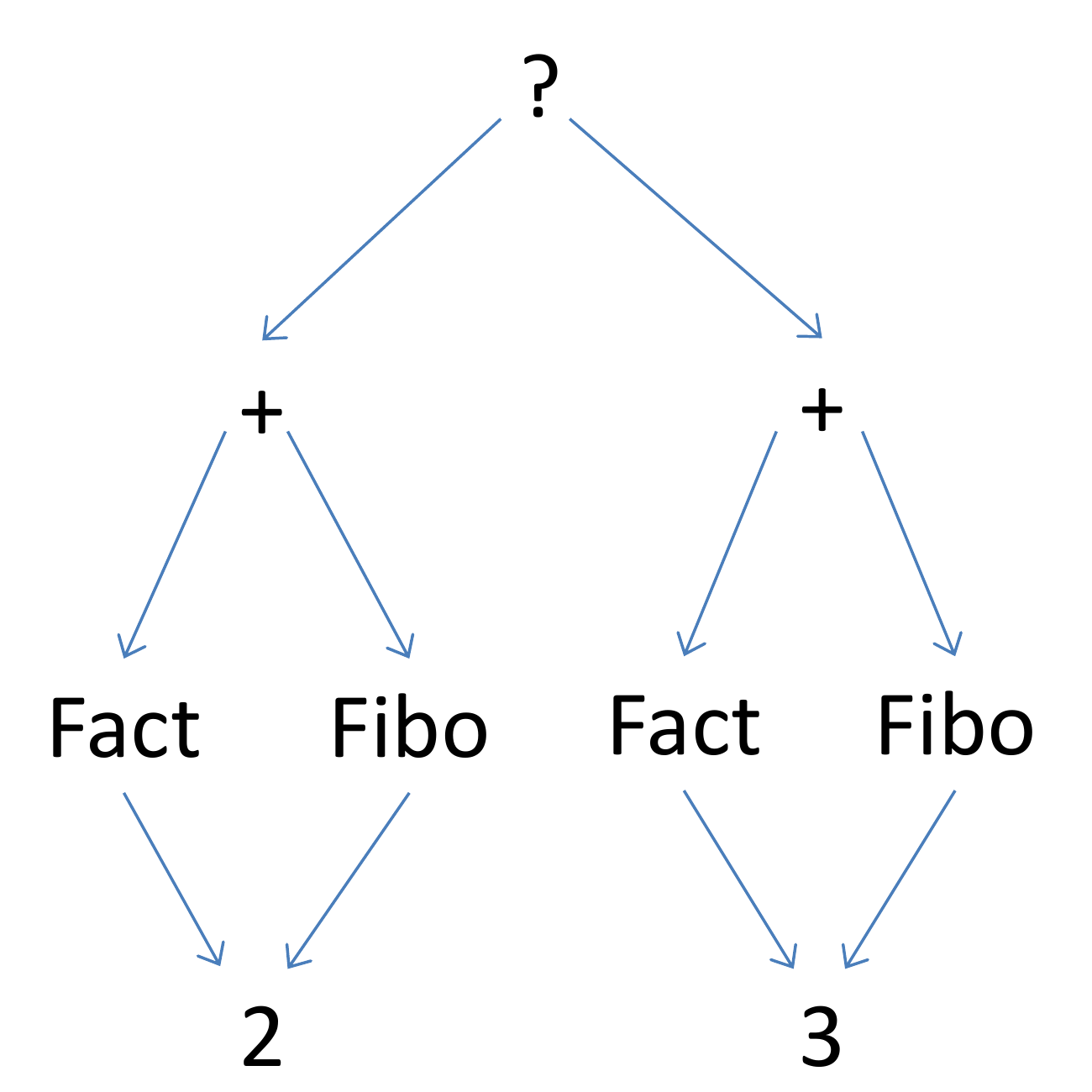}
\caption{Bubbling transformation of the expression in equation \ref{BubbleExampleEq}}
\label{fig:BubblingExB}
\end{minipage}
\end{figure} 

{Figure \ref{fig:BubblingExA} is a term graph representation of the expression in 	equation \ref{BubbleExampleEq}, where {\bf Fact} and {\bf Fibo} represent the unary defined operations factorial and fibonacci respectively. Figure \ref{fig:BubblingExB} represents a bubbling transformation of the term graph represented in Figure \ref{fig:BubblingExA} where the root of the graph, the node labeled {\bf +}, is the dominator of the choice operator and the ancestral path consists of the nodes labeled {\bf Fact}, {\bf Fibo} and {\bf +}. The choice operator becomes the root of the transformed graph and all necessary updates of the graph components are incorporated in the transformed graph. Details of all the updates are to follow.}

{A dominator of a node x is a node that appears in all the paths that connect the node x to the root of the graph. Therefore, the root of the graph is a dominator for all nodes of the graph other than itself. All the nodes that connect a node x to its dominator are part of the ancestral path. Our goal is to find a ``least dominator'' for a node. By least, we mean that the dominator will be the nearest to the node in question than all other dominators to minimize the number of nodes that are part of the ancestral path. We will see the benefits of such approach after defining the bubbling function.}

{Bubbling: GRAPH $\times N_g \times N_g \times 2^{N_g} \to$ g}

{Bubbling ($\langle N_g$, $L_g$, $E_g$, $Root_g\rangle$, x, a, AP) = $\langle N_{g_a}$, $L_{g_a}$, $E_{g_a}$, $Root_{g_a}\rangle$}

{The bubbling function takes a graph g, a source node x, a dominator node a and the ancestral path AP that connects x to a. Bubbling transforms the original graph by copying all the nodes that are part of AP and ``bubbles'' the node x up in the graph to take the place of node a. All the edge links must be maintained as mentioned in the definition of Edge links below ($E_{g_a}$). The root of the graph may change when the dominator is $Root_g$.}

\begin{figure}[ht]
\centering
\includegraphics[scale=0.17]{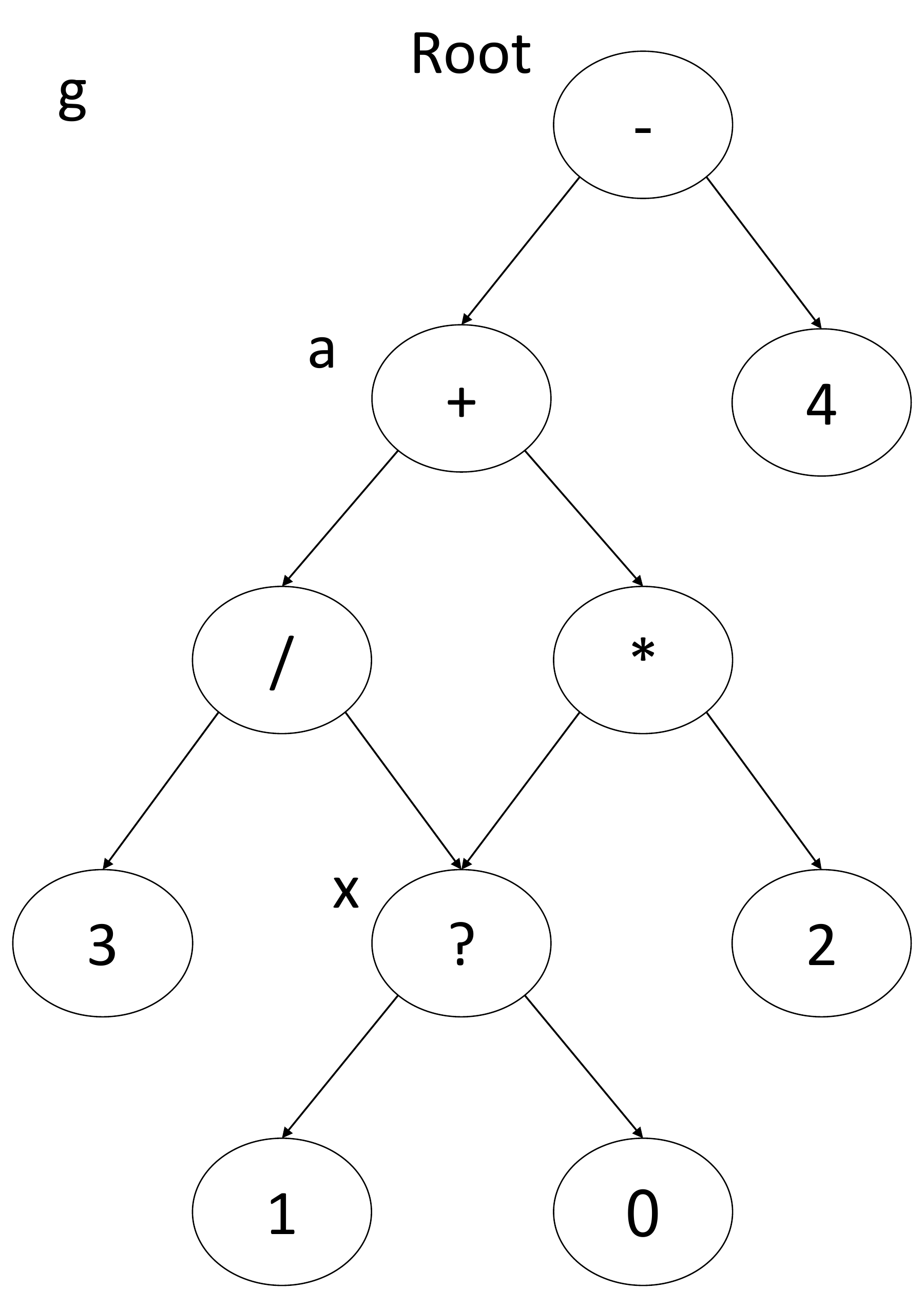}
\caption{Pictorial representation of equation \ref{bubblingEx}}
\label{fig:BubblingA}
\end{figure}

\begin{figure}
\centering
\begin{minipage}[ht]{0.4\linewidth}
\includegraphics[scale=0.18]{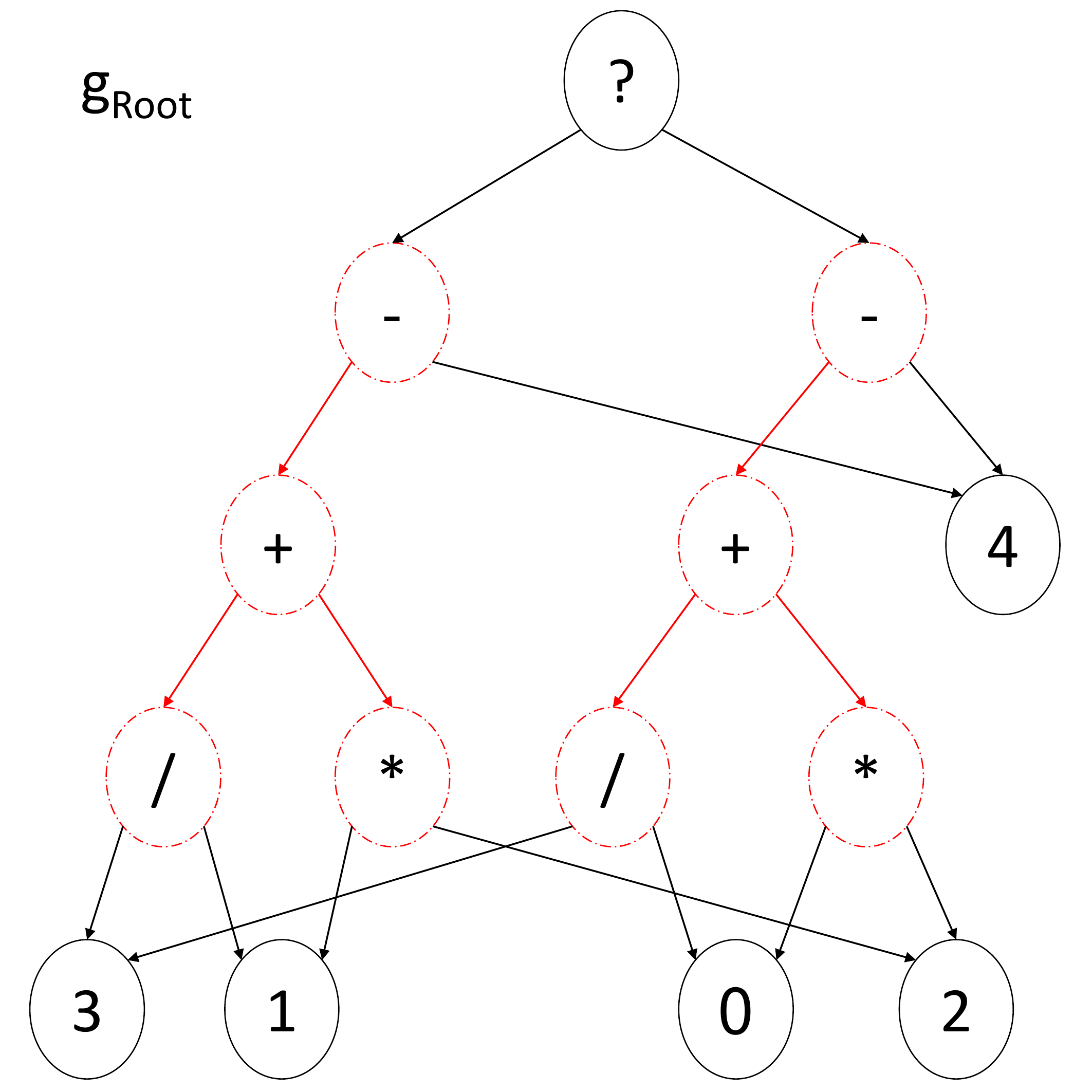}
\caption{Graph Bubbling with the root as dominator}
\label{fig:BubblingB}
\end{minipage}
\qquad
\begin{minipage}[ht]{0.4\linewidth}
\includegraphics[scale=0.18]{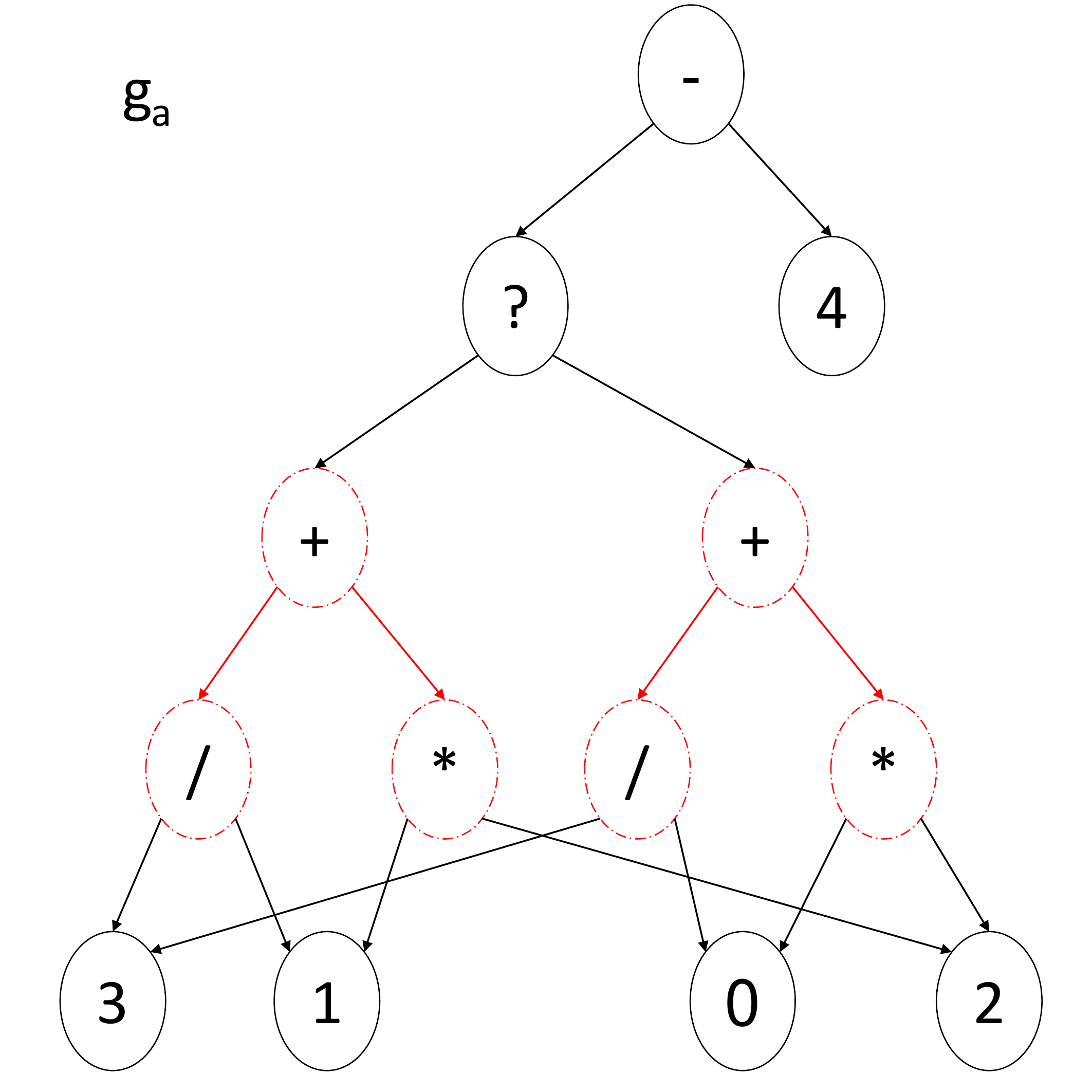}
\caption{Graph Bubbling with node a as dominator}
\label{fig:BubblingC}
\end{minipage}
\end{figure}

{Having several copies of the nodes that are part of AP motivates the need to find a dominator that minimizes such set. Figure \ref{fig:BubblingA} represents a pictorial representation of the expression in equation \ref{bubblingEx}. The transformation of the graph by bubbling with the dominator node a, represented in Figure \ref{fig:BubblingC}, copies less nodes than bubbling with the root as the dominator, represented in figure \ref{fig:BubblingB}. Please note that the copying approach discussed in section \ref{PrevAppSubSect} is a special case of bubbling where the dominator is the root of the graph.}

\equprogram{
\label{bubblingEx}
( ( (3 / X) + (X $\times$ 2) ) - 4) where X = ( 0 ? 1 )
}

{We describe here the transformations of the different components of the original graph after a call to the function bubbling:}
\newline
$N_{g_a}$ = ($N_g$ $\setminus$ AP ) $\cup A_i$ where $A_i$ =  $\displaystyle\bigcup_{i=1}^{arity(x)}n_i \forall$ n $\in$ AP $\wedge$ $N_g$ $\cap$ $A_i$ = $\emptyset$ 
\newline
$L_{g_a}$(n) = $L_g$(n) \quad if n $\notin$ AP
\newline
$L_{g_a}$($n_i$) = $L_g$(n) \quad if n $\in$ AP $\wedge$ 0 $<$ i $\leq$ arity(x)
\newline
\[
E_{g_a}(n, pos) = \left\{ 
\begin{array}{l l}
E_g(n, pos) & \quad \text{if n $\notin$ AP $\wedge$ n $\not=$ x}\\
x & \quad \text{if $E_g$(n, pos) = a}\\
a_{pos} & \quad \text{0 $<$ pos $\leq$ arity(x) $\wedge$ n = x}\\     
\end{array} \right.
\]
\newline
\[
	E_{g_a}(n_i, pos) = \left\{ 
\begin{array}{l l}
m_i & \quad \text{if n, m $\in$ AP $\wedge$ $E_g$(n, pos) = m}\\
m & \quad \text{if n $\in$ AP $\forall$ m $\notin$ AP $\wedge$ m $\not=$ x $\wedge$ $E_g$(n, pos) = m}\\
p & \quad \text{if n $\in$ AP $\wedge$ $E_g$(n, pos) = x $\wedge$ $E_g$(x, i) = p}\\     
\end{array} \right.
\]
\newline
\[
Root_{g_a} = \left\{ 
\begin{array}{l l}
Root_g & \quad \text{if $Root_g \not=$ a}\\
x & \quad \text{if $Root_g$ = a}\\     
\end{array} \right.
\]
\newline
\section{Implementation Details}\label{implementationDetailsSect}

{In this section we describe the details of the current stage of the implementation of the evaluation strategy used to compute functional logic language expressions and the bubbling transformation needed to model non-determinism in such expressions.}

\subsection{Symbols}\label{symbolsSubSect}

{Symbols represent elements of both $\Sigma$ and \vars. Each symbol has a name, an operator designation and a value. The name is represented as a string, while operator designation and value as integer values. The main operator designation represented are defined operations, data constructors and variables. We add also two special cases of operator designation: numbers and fail.  Numbers are a special case of data constructors separated from their main operator designation to speed up computations. Values are only associated with symbols of the operator designation number and are represented as integers. The fail symbol is also added to represent the inability to compute a $(head)$ $normal$ $form$ of the expression. All predefined symbols are kept in an array of distinct symbols to ensure no data constructor or defined operation has the same name.}

\subsection{Expressions}\label{expressionsSubSect}

{An expression is represented as a graph. We will represent the graph in a table. Each expression entry is represented by the following: (1) A root that represents the symbol of the expression. (2) An array of arguments that represent the positions of the arguments in the terms table. (3) An array of back pointers that contains the positions of the immediate parents of the expression in the table. (4) Minimum and maximum depth of the expression in the expressions table represented by how far an expression is from the root expression of the table. (5) A tag that represents whether a $normal$ $form$ of this expression has been reached or not.}
\begin{figure}
\centering
\begin{minipage}[ht]{0.4\linewidth}
\includegraphics[scale=0.14]{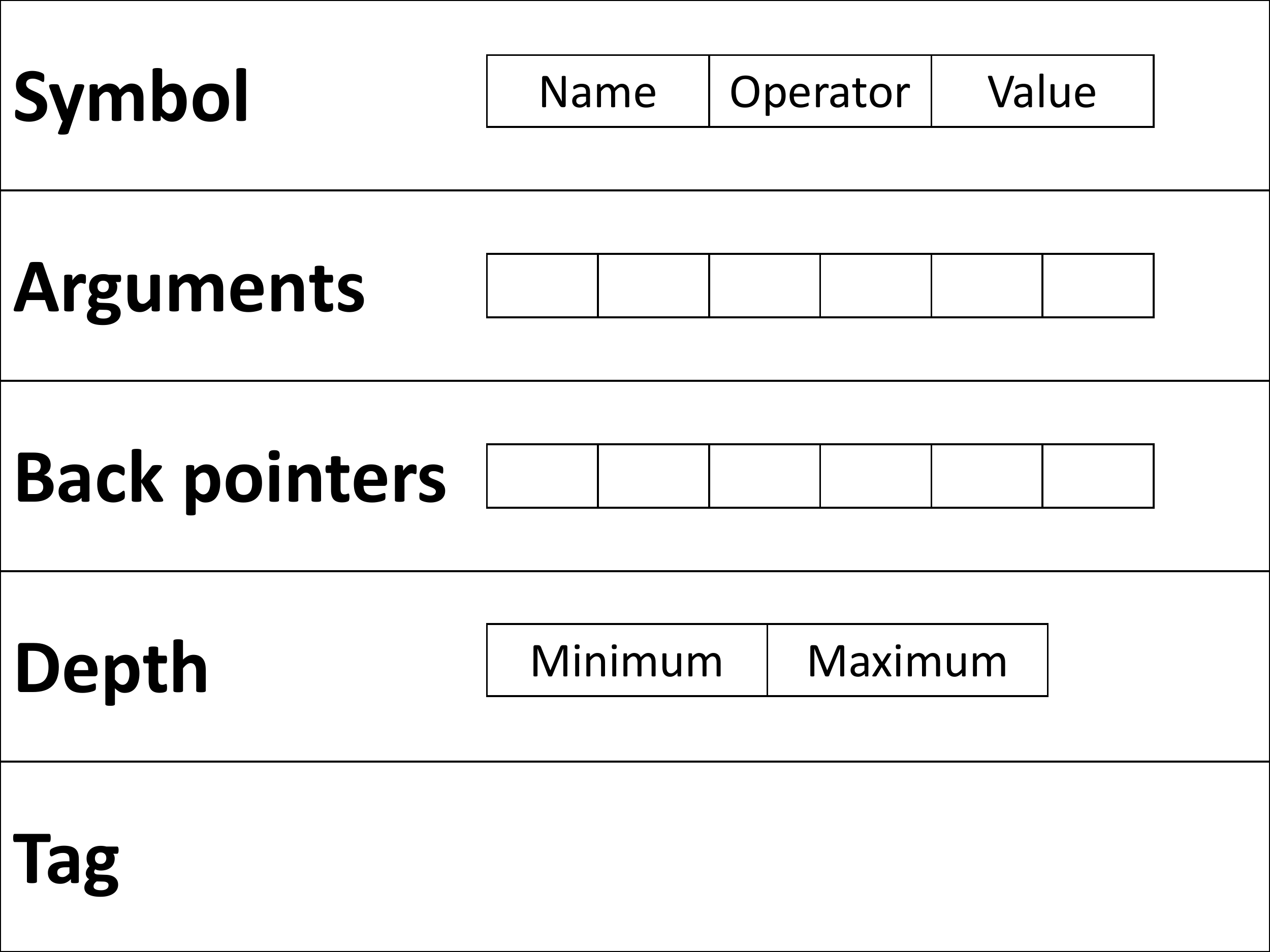}
\caption{Pictorial representation of a typical term entry in the expressions table}
\label{fig:3A}
\end{minipage}
\qquad
\begin{minipage}[ht]{0.4\linewidth}
\includegraphics[scale=0.14]{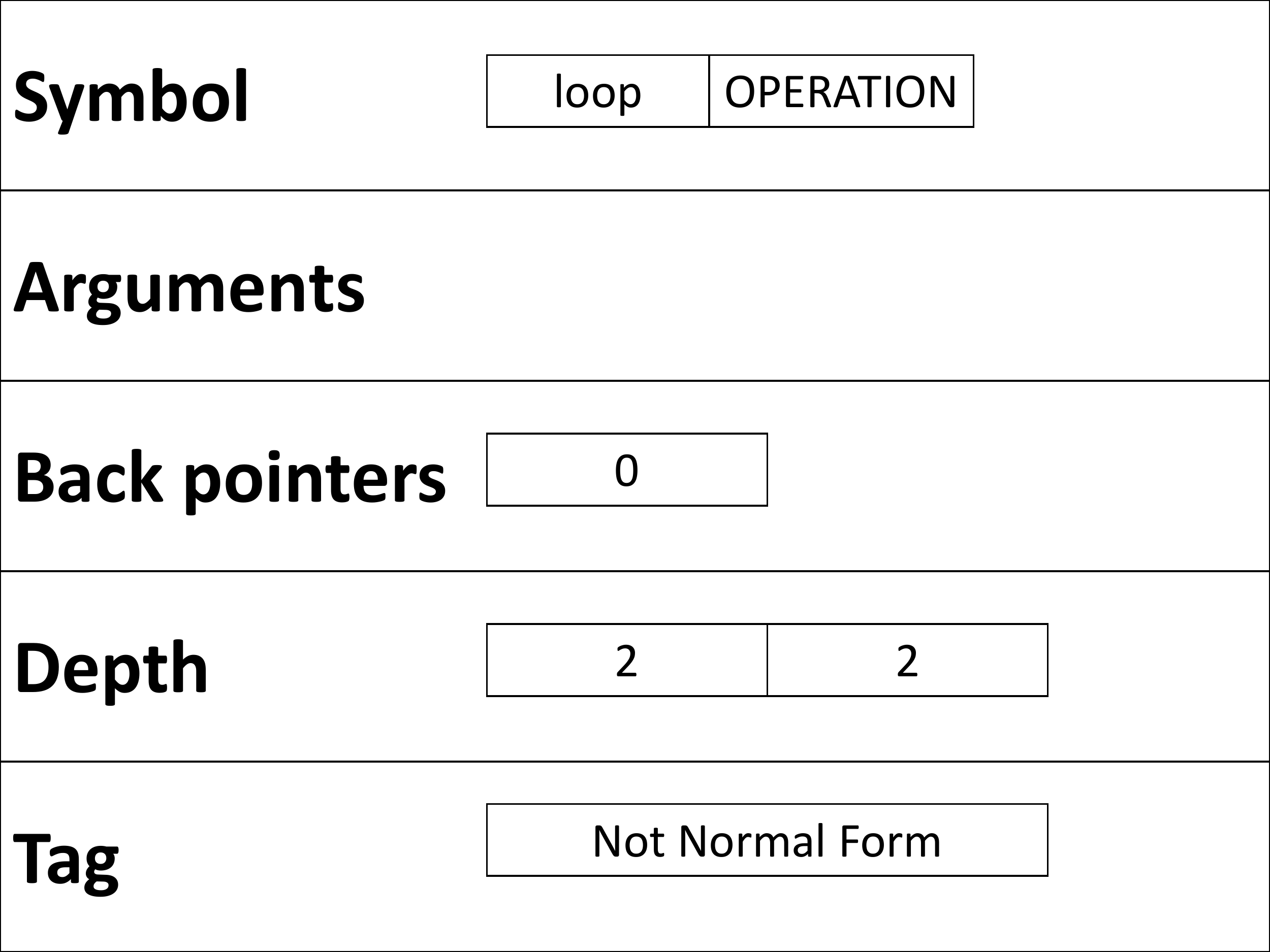}
\caption{Pictorial representation of the node labeled ``loop'' in the expression in equation \ref{loopEx}}
\label{fig:3B}
\end{minipage}
\end{figure}
\subsection{Parser}\label{parserSubSect}

{The first stage of computing an expression is to parse the expression needing evaluation by checking whether it is well-formed. The duty of the parser is to build a term graph and populate the expression table mentioned in section \ref{expressionsSubSect}. Expressions are built as directed acyclic graphs. Sharing of expressions is done through variables.}

\subsection{Rewrite rules}\label{rewriteRulesSubSect}

{Rewrite rules are coded using definitional trees. The definitional tree is built in this implementation manually. All the rewrite rules are coded in a single function and the program will branch to the appropriate rewrite rule according to the defined operation needing evaluation. When a defined operation t needs evaluation, the steps in Figure \ref{fig:rules} will take place to decide what is required to compute t.}
\begin{figure}
\begin{verbatim}
Switch (needed position n)
    case n is FAIL
        rewrite as FAIL
    case n is OPERATION
        step on n
    case n is CONS or NUMBER
        if n matches LEAF
            rewrite to right-hand-side
        else if n matches INNER BRANCH
            Continue on another needed position
        else
            rewrite as FAIL
        End if
End Switch
\end{verbatim}
\caption{function to match a needed position using Definitional Trees}
\label{fig:rules}
\end{figure}
\subsection{Computations}\label{ComputationsSubSect}

{The computation of an expression in the virtual machine is done by rewriting it to an expression in $head$ $normal$ $form$ or $normal$ $form$ using the rewrite rules.}

{The evaluation of an expression to an expression in $head$ $normal$ $form$ is done by rewriting $operation$-$rooted$ expressions using rewrite rules to a $constructor$-$rooted$ expression. All variables and $constructor$-$rooted$ expressions are already in $head$ $normal$ $form$ as discussed earlier in section \ref{PreliminariesSubSect}. The initial step taken to compute $head$ $normal$ $form$ for an expression is done by inspecting the root of the expression to be evaluated.}
\begin{figure}
\begin{verbatim}
Step (term t, mode m)
    if t is FAIL or NUMBER
        exit
    else if t is CONS
        if m is NORMAL FORM
             collect all OPERS ops part of t
             if any ops is FAIL
                 rewrite as FAIL
                 exit
             End If
             step on all ops
             exit
        else if m is HEAD NORMAL FORM
             exit
        End if
    else if t is OPERATION
        goto rewrite rule of t
        execute rewrite rule outcome
    End if
End Step
\end{verbatim}
\caption{Procedure to perform a ``step'' on a term}
\label{fig:Step}
\end{figure}
{Evaluating an expression to be in $normal$ $form$ is done through evaluating the expression to $head$ $normal$ $form$ and then checking that all of the expressions arguments are also in $normal$ $form$. This will guarantee that the expression and all its arguments are $constructor$-$rooted$ expressions containing no $operation$-$rooted$ expressions. The expression ``cons(2, nil)'' is in $normal$ $form$, while the expression ``cons((1 + 1), nil)'' is in $head$ $normal$ $form$ but not in $normal$ $form$.}

{The evaluation strategy will only inspect the needed arguments of the defined operation. A single rewriting step is executed on the arguments needing further evaluation or a rewriting of the defined operation will be done. A single computation step guarantees that the arguments are ``touched'' only once during a single round of evaluation. On the other hand, full evaluation of the arguments to $head$ $normal$ $form$ resembles the backtracking approach discussed in section \ref{PrevAppSubSect}, which is operationally incomplete. An example of a round of single ``rewriting step'' is: ({\bf Fact} 5 + {\bf Fact} 6) $\to$ ( ({\bf Fact} 4 $\times$ 5) + ({\bf Fact} 5 $\times$ 6)). Several rounds will be conducted until a $head$ $normal$ $form$ for the expression is reached.}

{The process of evaluating an expression to $normal$ $form$ will start by first evaluating the expression to $head$ $normal$ $form$, i.e., $constructor$-$rooted$ expression and then evaluating all the arguments of the $constructor$-$rooted$ expression to $normal$ $form$. If any of the arguments of the expression is of type ``fail'', failure will be declared and no $normal$ $form$ is obtained from such expression.}

\subsection{Non-determinism}\label{nonDeterminismSubSect4}

{Non-determinism represented by the choice operator is the main issue tackled by this implementation. When a choice is needed the steps in figure \ref{fig:Choice} will be performed.}

\begin{figure}
\begin{verbatim}
CodeChoice (term t)
    Remove all FAIL from arguments(t)
    if all arguments(t) are OPERATIONS
        step on arguments(t)
        exit
    else if any arguments(t) is in HEAD NORMAL FORM
        if arity(t) = 0
            rewrite t as FAIL
        else if arity(t) = 1
            rewrite t as argument(t)
        else if arity(t) > 1
            P = PathToRoot(t)
            dominator, AP = AncestralPath(P)
            Bubble(t, dominator, AP)
            exit
        End if
    End if
End CodeChoice
\end{verbatim}
\caption{Procedure to perform a ``rewrite step'' on a choice term}
\label{fig:Choice}
\end{figure}

{The calculation of the ``least dominator'' and ancestral path AP of the choice node is done by finding the farthest node from the root of the graph that is a dominator of the choice node. Finding the ``least dominator'' is done with the aid of some heuristic information about the minimum and maximum depth range of each node in the graph. The minimum depth represents the shortest distance this node can be from the root of the graph, while the maximum depth represents the farthest distance this node can be from the root of the graph. The depth of a node can change by adding a parent to such node or by a change in the depth of a current parent. To minimize the effect of updating the minimum and maximum depths of a node, the minimum depth will be decreased in case the new minimum depth is smaller than the current minimum depth. The maximum depth is updated if the new maximum depth is greater than the old maximum depth. Deleting a parent may cause a change in the minimum and maximum depths of the node but the virtual machine will not update the depths of the node because of the high cost incurred from such update. This will only affect the efficiency of finding the ``least dominator'' of a node.}

\section{Future work}\label{FutureWorkSect}

{The current implementation can successfully compute $normal$ $forms$ of expressions containing deterministic and non-deterministic operations. The evaluation strategy used keeps making single computation steps on the root of the expression until a $normal$ $form$ is reached. A step on a deterministic operation either rewrites the expression when all its needed arguments are in $normal$ $form$ or performs a step on each needed argument not in $normal$ $form$.}
\newline
{The current implementation lacks an implementation of narrowing or logical variables. The presence of an efficient implementation of narrowing is extremely important. The essence of functional logic computations is to incorporate a mechanism for dealing with missing and partial information, which is satisfied by logical variables.}
\newline
{The long-term goal of the implementation is exploitation of parallelism. According to Gupta et al. declarative and logic programming is seen to be well suited for parallelism \cite{parallelprolog}. Exploitation of and-parallelism, or-parallelism or both is possible. And-parallelism can be exploited by collecting and executing several needed expressions at the same time. However, synchronization of expressions may be needed to express dependencies based on sharing. Or-parallelism can be exploited by trying different arguments of the choice simultaneously.}
\newline
{Further tuning of the evaluation strategy is needed. The evaluation strategy adopted in the current implementation exhibits a naive implementation of breadth-first evaluation and needs additional improvements. The automatic generation of code for definitional trees is a desirable feature to be implemented in the future.}
\newline
{Finding the ``least dominator'' needs an efficient implementation. The literature has some good results for ``offline'' graph settings where queries can be answered in constant time, but no current efficient algorithm for a dynamic graph is yet available.}
 
\section{Conclusion}\label{conclusionSect}

{We introduced an implementation of bubbling, a mechanism that tackles non-determinism correctly and efficiently. Backtracking is incomplete and cannot model non-determinism. On the other hand, Copying is a special case of bubbling where the root is always considered the dominator. This approach is complete but lacks efficiency due the reconstruction of nodes unnecessarily if failure is detected in one of the alternatives of a choice in an early stages.}

\bibliographystyle{plain}
\bibliography{abdulla_bib}

\end{document}